%Paper: 9112031
%From: LEAF@HUHEPL.HARVARD.EDU
%Date: Fri, 13 Dec 1991 16:15:56 -0500 (EST)

%%%%%%%%%%%%%%%%%%%%%%%%%%%%%%%%%%%%%%%%%%%%%%%%%%%%%%%%%%%%%%%%%%%%%%%%%%%%
%%%%%%%%%%%%%%%%%%%%%%%%%%%%%%%%%%%%%%%%%%%%%%%%%%%%%%%%%%%%%%%%%%%%%%%%%%%%
%%%%%       Please Note:  This file has eight figures appended         %%%%%
%%%%%       at the end, beneath the statement `cut here for            %%%%%
%%%%%       figures.ps.'  You should detach the file below the         %%%%%
%%%%%       cut line and print the figures separately as a             %%%%%
%%%%%       PS file.                                                   %%%%%
%%%%%%%%%%%%%%%%%%%%%%%%%%%%%%%%%%%%%%%%%%%%%%%%%%%%%%%%%%%%%%%%%%%%%%%%%%%%
%%%%%%%%%%%%%%%%%%%%%%%%%%%%%%%%%%%%%%%%%%%%%%%%%%%%%%%%%%%%%%%%%%%%%%%%%%%%
\input harvmac
\pretolerance=750
\let\<=\langle
\let\>=\rangle
\let\ol=\bar
\let\ov=\over
\let\p=\partial
\let\ov=\over

\let\ka=\kappa
\let\de=\delta
\let\De=\Delta
\let\ga=\gamma
\let\Ga=\Gamma
\let\th=\theta
\let\rh=\rho
\let\ph=\varphi

\let\la=\lambda
\let\th=\theta
\let\l=\left
\let\r=\right
\let\what=\widehat
\def\e{\hbox{e}}
\def\Yt{\tilde Y}

\Title{\vbox{\baselineskip12pt\hbox{HUTP-91/A061}}}
{\vbox{\centerline{From Here to Criticality:}
\vskip2pt\centerline{Renormalization Group Flow}
\vskip2pt\centerline{Between Two Conformal Field Theories}}}

\centerline{W. A. Leaf-Herrmann\footnote{$^\dagger$}
{(leaf@huhepl.hepnet, \ @huhepl.bitnet, \ or \ @huhepl.harvard.edu)}}
\bigskip\centerline{Jefferson Physical Laboratory}
\centerline{Harvard University}\centerline{Cambridge, MA 02138}

\vskip .3in

Using nonperturbative techniques, we study the renormalization group
trajectory between two conformal field theories.
Specifically, we investigate a perturbation of the $A_3$ superconformal
minimal model such that in the infrared limit the theory flows to the $A_2$
model.
The correlation functions in the topological sector of the theory are computed
numerically along the trajectory, and these results are compared to the
expected asymptotic behavior.
Excellent agreement is found, and the characteristic features of the infrared
theory, including the central charge and the normalized operator product
expansion coefficients are obtained.
We also review and discuss some aspects of the geometrical description of
$N=2$ supersymmetric quantum field theories recently uncovered by S. Cecotti
and C. Vafa.

\Date{12/91}

\newsec{Introduction}

The set of conformal field theories are the infrared or ultraviolet fixed
points, or critical points, of renormalization group flow trajectories in the
space of two-dimensional quantum field theories.
At these fixed points, conformal invariance provides a set of constraints,
realized by the infinite-dimensional Virasoro algebra, which often allow for
the exact solution of the theory, in principle via the BPZ bootstrap \ref
\rBPZ{A.A. Belavin, A.M. Polyakov, and A.B. Zamolodchikov, Nucl. Phys. B241
(1984) 333.}, or in
practice via explicit representations of the Virasoro algebra \ref
\rGKO{P. Goddard, A. Kent, and D. Olive, Commun. Math. Phys. 103 (1986) 105.}.
However, these methods generically allow us to solve the theory only at the
critical point.

To better understand the structure of the space of two-dimensional field
theories, and the special role played by the conformal field theories, we
would like to be able to compute the correlation functions both on and off of
the critical point.
Typically, the best that can be done is to use conformal perturbation theory
in the neighborhood of a fixed point, as demonstrated by Zamolodchikov \ref
\rZ{A.B. Zamolodchikov, Sov. J. Nucl. Phys. 46 (1987) 1090.} and
others \nref\rLC{A.W.W. Ludwig and J.L. Cardy, Nucl. Phys. B285 [FS19] (1987)
687.}\nref
\rP{R.G. Pogosyan, Sov. J. Nucl. Phys. 48 (1988) 763.}\nref
\rKMS{D.A. Kastor, E.J. Martinec, and S.H. Shenker, Nucl. Phys. B316 (1989)
590.}\nref
\rLH{W.A. Leaf-Herrmann, Nucl. Phys. B348 (1991) 525.}\nref
\rBM{J. Bie\'{n}kowska and E. Martinec, {\it The renormalization group flow in
$2D$ $N=2$ SUSY Landau-Ginsburg models}, Enrico Fermi Institute preprint EFI
90-62.}
\refs{\rLC{--}\rBM}.
For example, one would like to be able to calculate the scaling behavior of
the quantum fields in a conformal theory perturbed by some relevant field,
under the action of renormalization group flow.
The infrared limit of this theory should correspond to some (possibly
trivial) conformal field theory.
While this question can be addressed perturbatively in some cases, for
instance the minimal models near $c=1$ \rZ, we generally require some
nonperturbative techniques to answer the above question.

\nref\rCGPi{S. Cecotti, L. Girardello, and A. Pasquinucci, Nucl. Phys. B328
(1989) 701.}
\nref\rCGPii{S. Cecotti, L. Girardello, and A. Pasquinucci, Int. J. Mod. Phys.
A6 (1991) 2427.}
\nref\rSCi{S. Cecotti, Int. J. Mod. Phys. A6 (1991) 1749.}
\nref\rSCii{S. Cecotti, Nucl. Phys. B355 (1991) 755.}
\nref\rCV{S. Cecotti and C. Vafa, {\it Topological Anti-Topological Fusion},
Harvard preprint HUTP-91/A031, SISSA-69/91/EP, to appear in Nucl. Phys.}
In the case of two-dimensional field theories with $N=2$ supersymmetry some of
the requisite techniques have recently been developed \refs{\rCGPi{--}\rCV},
which allow for the nonperturbative calculation of a class of correlation
functions both on and off the critical point.
This class of correlation functions is known as the topological sector of
$N=2$ field theories, and is closely related to the correlation functions of
physical
observables in topological quantum field theories
\ref\rW{E. Witten, Commun. Math. Phys. 118 (1988) 411.}.
The topological sector is composed of the expectation values of chiral fields
evaluated between the set of supersymmetric ground states.

\nref\rG{D. Gepner, Nucl. Phys. B296 (1987) 757.}
\nref\rEM{E.J. Martinec, Phys. Lett. 217B (1989) 431.}
\nref\rVW{C. Vafa and N.P. Warner, Phys. Lett. 218B (1989) 51.}
\nref\rGVW{B. Greene, C. Vafa, and N.P. Warner, Nucl. Phys. B324 (1989)
371.}
\nref\rM{E. Martinec, {\it Criticality, Catastrophes, and Compactifications},
V.G. Knizhnik memorial volume, 1989.}
\nref\rNS{N. Seiberg, Nucl. Phys. B303 (1988) 286.}
\nref\rDKL{L. Dixon, V. Kaplunovsky, and J.Louis, Nucl. Phys. B329 (1990) 27.}
The equivalence between two-dimensional $\sigma$-models on Calabi-Yau spaces
and certain $N=2$ superconformal models, first observed by Gepner \rG, is
well-known, and $N=2$ Landau-
Ginsburg effective Lagrangians provide explicit realizations of this
correspondence \refs{\rEM{--}\rM}.
This equivalence has led to the application of geometrical methods in the
characterization of $N=2$ superconformal field theories
\refs{\rCGPii,\rNS,\rDKL}.
Using the quasi-topological nature of $N=2$ supersymmetry, S. Cecotti and C.
Vafa have recently uncovered the generalization of these geometrical aspects
to arbitrary $N=2$ quantum field theories \refs{\rSCii,\rCV}.
In this paper we shall apply their results to the nonperturbative calculation
of the topological sector of a theory which interpolates between two conformal
field theories along the renormalization group trajectory connecting them.

In Section 2 we define and discuss the properties of the topological sector of
$N=2$ supersymmetric quantum field theories, and the relation between this
sector and topological quantum field theories based on twisted $N=2$ models.
The basic geometrical framework needed to solve for the topological sector
correlation functions is reviewed in Section 3.
Section 4 describes how, by using the $N=2$ non-renormalization theorem,
these correlation functions may be calculated nonperturbatively along a
renormalization group trajectory.
We also discuss two quantities introduced in \rCV\ which serve to characterize
the theory both at the conformal point, and off of criticality, known as the
Ramond charge matrix and the algebraic $Q$-matrix.
Both are easily computed in the topological sector.
As a concrete application of this framework, we analyze the renormalization
group flow between two conformal field theories, the $A_3$ $N=2$ minimal model
perturbed in such a way that the theory flows to the $A_2$ model in the
infrared limit.
The computation is discussed in Section 5, and we compare the results of the
nonperturbative numerical calculation of the correlation functions of the
interpolating theory with the expected asymptotic behavior in Section 6.
Our conclusions are presented in Section 7.

\newsec{Topological Sector of $N=2$ Quantum Field Theories}

We consider two-dimensional quantum field theories with an $N=2$
supersymmetry.
We assume the topology of the two-dimensional space to be a cylinder, or
equivalently, a sphere with two punctures.
The supercharges $Q^+$, $Q^-$, and their Hermitian conjugates $\ol
Q^-$, $\ol Q^+$, obey the algebra:
\eqn\eSuAl{\Bigl\{Q^+,Q^-\Bigr\}=-\p,\qquad\qquad \l\{\ol Q^+,\ol
Q^-\r\}=-\ol\p,}
with all other (anti-)commutation relations vanishing.
We impose periodic boundary conditions on the fermionic operators, so that
the Witten index, $Tr\l(-1\r)^F$, where $F$ is the operator which counts
fermion
number, is well-defined.
Hence we restrict to the Ramond sector.
We shall restrict our study to theories in which supersymmetry is not
spontaneously
broken, so we shall assume that there are $\De$ supersymmetric ground states
$|i\>$ satisfying
\eqn\eSVac{Q^\pm|i\> = \ol Q^\pm|i\> = H|i\> = 0, \quad 0\le i \le \De-1,}
where $H$ is the Hamiltonian of the theory.
In the special case of $N=2$ Landau-Ginsburg theory $Tr\l(-1\r)^F = \De$.

The set of chiral superfields are those superfields $X_i$ which satisfy
\eqn\eCH{\Bigl[Q^+,X_i\Bigr] = \l[\ol Q^+,X_i\r] = 0.}
It follows as an immediate consequence of \eSuAl\ and \eCH\ that
\eqna\eDCH
$$\eqalignno{-\p X_i &= \Bigl\{ Q^+,\Bigl[Q^-,X_i\Bigr]\Bigr\}&\eDCH a\cr
-\ol\p X_i &= \l\{ \ol Q^+,\l[\ol Q^-,X_i\r]\r\}.&\eDCH b\cr}$$
Together these relations have an important consequence for the set of Green
functions involving chiral fields and the supersymmetric ground states.
They are independent of the positions of the chiral fields:
\eqn\eDGz{\eqalign{-{\p\ov{\p z_a}}\<\ol j|\cdots X_k\l(z_a\r) \cdots |i\> &=
\<\ol j|\cdots \l\{ Q^+,\l[Q^-,X_k\l(z_a\r)\r]\r\} \cdots |i\>\cr
&=0,\cr}}
since $Q^+$ commutes with all the chiral fields and annihilates the ground
states.
This set of Green functions, composed of matrix elements involving chiral
fields evaluated between supersymmetric ground states is known as the
topological sector of $N=2$ supersymmetric theories.

Since the above Green functions are independent of the positions of the chiral
fields, it is apparent that the product of two chiral fields contains no short
distance singularities.
By defining the point-wise product of chiral fields as
\eqn\ePWP{X_iX_j\l(z_a\r) = \lim_{z_b\to z_a}\ X_i\l(z_a\r)X_j\l(z_b\r),}
we see that the set of chiral fields forms a commutative ring ${\cal R}$,
since
the
product of any two chiral fields is also chiral.
In the context of the critical theory, this chiral ring essentially
characterizes the conformal field theory \ref\rLVW{W. Lerche, C. Vafa, and N.
Warner, Nucl. Phys. B324 (1989) 427.}.
We can extend the notion of a product of chiral fields to chiral fields which
are located at different points using an equivalence relation.
If we choose a basis for the set of chiral fields, say $\{\ph_i\}$, then we
have
\eqn\eEQV{\ph_i\ph_j \sim C_{ij}{}^k\ph_k,}
where the equivalence is modulo $Q^+$ and $\ol Q^+$ commutator terms, since by
\eDCH\ above the difference between a chiral field evaluated at two different
points is terms of this form.

Another important fact about $N=2$ quantum field theories is that the chiral
ring ${\cal R}$ is isomorphic to the vector space of supersymmetric ground
states as ${\cal R}$-modules.
This isomorphism is the spectral flow \ref\rSS{A. Schwimmer and N.
Seiberg, Phys. Lett. B184 (1987) 191.}.
Hence we may identify the supersymmetric ground states by the operation of a
chiral field on the unique ground state, $|0\>$, which is the image under
spectral flow of the
identity operator in the chiral ring:
\eqn\eBGS{|i\> \sim  \ph_i|0\>.}
Using the relation \eEQV\ for the chiral fields, we then have the
following equivalence relation for chiral fields acting on the ground states:
\eqn\eEQVV{\ph_i|j\> \sim C_{ij}{}^k|k\>,}
where this equivalence is modulo $Q^+$ and $\ol Q^+$ acting on some state.

Using the above relations, we can now reduce the calculation of any Green
function in the topological sector to essentially matrix multiplication.
For example, if we wish to compute the correlation function
$$\<\ol j| \ph_m \ph_n |i\>,$$
then using \eEQVV, and the fact that any states which are $Q^+$- (or $\ol Q^+
$-)exact annihilate $\<\ol j|$, we find that this computation reduces to the
product
\eqn\eCGF{\<\ol j| \ph_m \ph_n |i\> = g_{l\ol j} C_{mk}{}^l C_{ni}{}^k,}
where we define $g_{i\ol j}$ to be the Hermitian inner product for the
supersymmetric ground states:
\eqn\eDEFG{g_{i\ol j} = \<\ol j|i\>.}
Hence the calculation of the Green functions in the topological sector can be
reduced to the problem of calculating the $C_{ij}{}^j$ and the inner product
$g_{i\ol j}$ of the $N=2$ quantum field theory.

The above topological sector is closely related to the set of correlation
functions of physical observables in topological quantum field theory.
For every $N=2$ quantum field theory, we can define a topological quantum
field theory by twisting the energy-momentum tensor using the conserved
$R$-current $J_\mu$ corresponding to fermion number:
\eqn\eTST{T_{\mu\nu}\rightarrow T_{\mu\nu} + {1\ov4}\l(\p_\mu J_\nu +
\p_\nu J_\mu\r).}
This is equivalent to redefining the coupling of the theory to two-dimensional
gravity using the fermion numbers of the fields.
We then define a nilpotent BRST operator:
\eqn\eBRST{Q_{BRST} = Q^+ + \ol Q^+.}
The physical states of the topological theory are then defined to be the BRST
cohomology classes of this operator.
The fields which commute with the BRST operator are precisely the chiral
fields discussed above.

However, there is an important difference between topological quantum field
theories and the topological sector of an $N=2$ quantum field theory.
In topological theories, the natural inner product is defined by:
\eqn\eTOPIN{\eta_{ij} = \<j|i\> = \<0| \ph_j \ph_i |0\>.}
This symmetric inner product is truly a topological object, in the sense that
it is independent of the representative of the BRST cohomology class chosen to
define a state.
The Hermitian inner product $g_{i\ol j}$ does not share this property.
This is because in $N=2$ quantum field theory we define the adjoint of a state
by
\eqn\eHAD{\l(|i\>\r)^\dagger = \<\ol i| \sim \<\ol0|\ol\ph_{\ol i},}
where now the equivalence relation is modulo states which are $Q^-$- and
$\ol Q^-$-exact.
Since states created by applying chiral fields, $\ph_i$, to the state $|0\>$
are not annihilated by $Q^-$ or $\ol Q^-$, the inner product $g_{i\ol j}$ is
defined using precisely the supersymmetric ground states.
We could have defined a topological theory using the adjoint of the above BRST
operator \eBRST, and this would have produced a topological theory with the
set of physical observables being represented by the anti-chiral fields, those
superfields annihilated by $Q^-$ and $\ol Q^-$, instead.
S. Cecotti and C. Vafa have described this construction as an {\it
anti}-topological
theory, thus leading to the notion of the topological sector of $N=2$ field
theory as being the {\it fusion} of a topological theory and an
anti-topological
theory.

Since the bases $|i\>$ and $|\ol j\>$ correspond to two different set of
labels for the set of supersymmetric ground states, we can derive a useful
relation between $g_{i\ol j}$ and $\eta_{ij}$.
We define $M$ to be the complex matrix which relates these two bases:
\eqn\eDEFM{\<\ol i| = \<j|M_{\ol i}{}^j.}
$M$ is known as the real structure.
Using this definition of $M$, we find the following relationship between
$g_{i\ol j}$ and $\eta_{ij}$:
\eqn\eRS{g_{i\ol j} = \eta_{ik}M_{\ol j}{}^k.}
The CPT-invariance of the quantum field theory implies that $M$ satisfies
\eqn\eMM{MM^\ast = 1,}
and this leads immediately to the following condition:
\eqn\eRCON{\eta^{-1}g\l(\eta^{-1}g\r)^\ast = 1.}
This relation shall be referred to as the reality constraint.

\newsec{$N=2$ Landau-Ginsburg Models}

A particularly interesting class of $N=2$ supersymmetric quantum field
theories are those which can be represented by Landau-Ginsburg models.
These models are described by a Lagrangian density defined in terms of the
chiral (and anti-chiral) superfields of the theory by
\eqn\eLGL{{\cal L} = \half\int d\/^4\th\ K\l(X_i,\ol X_{\ol i}\r) + \l(\int
d\/^2
\th\ W\l(X_i\r) + h.c.\r),}
where the first term, which is integrated over all superspace, is known as a
$D$-term and includes the kinetic terms of the action, and the second term,
known as an $F$-term, is a holomorphic function of the chiral superfields.
This function, $W\l(X_i\r)$, is the superpotential of the theory.
In the case of Landau-Ginsburg models we shall find that the correlation
functions in the topological sector can be computed nonperturbatively, and
depend only on the superpotential.
The computation of the operator product expansion coefficients $C_{ij}{}^k$
and the
topological inner product $\eta_{ij}$ is quite easy in these models, and only
the calculation of $g_{i\ol j}$ is nontrivial.

The Green functions in the topological sector are independent of the kinetic
term, $K\l(X_i,\ol X_{\ol i}\r)$, essentially because due to the integral over
all
superspace, this term is $Q^+$-exact.
Hence the effect of any variation of the kinetic term on the correlation
functions in the topological sector will vanish:
\eqn\eVARK{\eqalign{\de\<\ol j| \ph_k |i\> &= -\int d\/^2z\,d\/^4\th\ \<\ol j|
\ph_k \de K\l(X_l,\ol X_{\ol l}\r)\l(z, \ol z\r)|i\>\cr
&= -\int d\/^2z\ \<\ol j| \ph_k\,\{Q^+,\Psi\l(X_l,\ol X_{\ol l}\r)\}\l(z,\ol
z\r)|i\>\cr
&= 0,\cr}}
where we have used the fact that $Q^+$ commutes with the chiral field $\ph_k$
and annihilates the supersymmetric ground states.
This formal argument can be made rigorous \rCGPii, but it establishes the
result that
the topological sector is essentially determined by the superpotential
$W\l(X_i\r)$
alone, and does not depend on the details of the kinetic term.

The superpotential also simply encodes the behavior of the ring of chiral
fields.
Using the Lagrangian \eLGL, we derive the following equation of motion for the
chiral fields:
\eqn\eEOM{\p_i W\l(X_j\r) = -\l\{Q^+,\l[\ol Q^+,\p_i K\l(X_j,\ol X_{\ol
j}\r)\r]\r\},}
and this defines which products of chiral fields are $Q^+$-exact in the
theory.
We may then define the chiral ring as
\eqn\eDCR{{\cal R} = {\bf C}\l[X_i\r]/\p_jW.}
The computation of the operator product coefficients $C_{ij}{}^k$, as defined
in \eEQV, is simply a matter of polynomial multiplication modulo the ideal
generated by $\{\p_iW\}$, which set the products of chiral
fields equal to zero if they are $Q^+$-exact,
\eqn\eEOMC{\p_i W\l(X_j\r) \sim 0.}
In particular, if one chooses a holomorphic basis for the chiral ring, i.e.
one that depends holomorphically on the complex parameters in the
superpotential, then the operator product coefficients will also be
holomorphic functions of these parameters.

The topological inner product $\eta_{ij}$ is also easily computed using a
result from the study of topological Landau-Ginsburg models.
It can be shown \ref
\rTLG{C. Vafa, Mod. Phys. Lett. A6 (1991) 281.}\ that
\eqn\eTLG{\eqalign{\eta_{ij} &= \<0|\ph_j \ph_i|0\>\cr
&= Res_W\l[\ph_j \ph_i\r],\cr}}
where $Res_W\l[F\l(X_k\r)\r]$ is defined using the Grothendieck residue by
\eqn\eGR{Res_W\l[F\l(X_k\r)\r] = {1\ov \l(2\pi i\r)^n} \int
dX_1\wedge\cdots\wedge
dX_n
\ F\l(X_k\r)\,\l(\p_1W\cdots\p_nW\r)^{-1},}
where $n$ is the number of chiral superfields in the theory.
In particular, one can show that the fundamental nonvanishing topological
correlation function is given by
\eqn\eRES{Res_W\l[H\r] = \mu,}
where $H$ is the Hessian of the superpotential $W$, defined by
\eqn\eHESS{H = \det\l(\p_i\p_j W\r),}
and $\mu$ is the criticality index of $W$, which is the same as the number of
supersymmetric ground states of the theory, or equivalently, the number of
elements in the chiral ring ${\cal R}$.
All other correlation functions $\<0|F\l(X_k\r)|0\>$ will vanish unless
$F\l(X_k\r)$
contains some scalar multiple of the Hessian $H$, modulo the ideal generated
by $\{\p_iW\l(X_j\r)\}$.
Again, in a holomorphic basis of the chiral ring ${\cal R}$, the inner product
$\eta_{ij}$ will only depend holomorphically on the parameters in the
superpotential.

When the superpotential $W\l(X_i\r)$ is a quasi-homogeneous polynomial of the
chiral
superfields, then it characterizes a conformal field theory
\refs{\rGVW,\rM,\rLVW}.
This statement relies on assumption that the usual non-renormalization
theorems for $N=2$ supersymmetry, for which perturbative proofs exist, hold
nonperturbatively as well.
In the case of such quasi-homogeneous superpotentials, it is known how to
directly compute the Hermitian inner product $g_{i\ol j}$ using the
superpotential.
This in itself is a very interesting result, since it allows one to compute
those correlation functions in the topological sector of the conformal theory
without an explicit representation of the Virasoro algebra.
The result for the inner product of two relevant chiral fields is given by
\refs{\rSCii,\rCV}\
\eqn\eGIJ{g_{i\ol j}=\int \prod_{l=0}^n dX_l\,d\ol X_{\ol l}\
\ph_i\l(X_k\r)\,\ol\ph_{\ol j}(\ol X_{\ol k})\,\exp\l(W-\ol W\r),}
where $\ph_i\l(X_k\r)$ is the element of the chiral ring which corresponds to
the
spectral flow of the ground state $|i\>$ and $n$ is the number of chiral
superfields in the theory.

To summarize the situation, for Landau-Ginsburg models we can easily compute
the operator product coefficient $C_{ij}{}^k$ as well as the topological inner
product $\eta_{ij}$, and in an appropriate basis they are holomorphic
functions of the parameters in the superpotential $W\l(X_i\r)$.
Furthermore, at the critical point, where the superpotential is quasi-
homogeneous, we can calculate the Hermitian inner product $g_{i\ol j}$.
To address the question of calculating the Green functions of the topological
sector we need to know how $g_{i\ol j}$ depends on the parameters in the
superpotential.
Thus we must determine how the supersymmetric ground states $|i\>$ depend on
the parameters in the superpotential.
This is the question to which we now turn.

Let us start with some superpotential, say one corresponding to a conformal
theory, for which we know the supersymmetric ground states $|i\>$.
We now consider adding perturbations to the superpotential using the elements
of the chiral ring ${\cal R}$:
\eqn\ePERL{\Delta{\cal L} = \int d\/^2\th\ t^i\ph_i\l(X_j\r) + h.c.,}
where $t^i$ are the perturbing couplings and the $\ph_i\l(X_j\r)$, defined by
\eqn\eDEFCH{\ph_i\l(X_j\r) = -{\p\ov\p t^i}W\l(X_j\r),}
are a basis for the chiral ring ${\cal R}$.
Geometrically, we may think of the parameters $t^i$ as complex coordinates on
the
space of supersymmetric deformations of the superpotential.
At each point in this space, we have an associated vector space defined by the
set of supersymmetric ground states.
Locally, the vacuum bundle is the product of these two spaces.
A natural connection $A_i$ may be defined on this bundle using the partial
derivatives of the ground states with respect to the parameters $t^i$ by:
\eqn\eDEFA{\p_i|j\> = A_{ij}{}^k|k\> + |\Psi\>,}
where $|\Psi\>$ is orthogonal to the space of ground states.
The components of the connection are defined by:
\eqn\eCOMPA{A_{ij\ol k} = \<\ol k|\p_i|j\> = A_{ij}{}^l g_{l\ol k}.}
Hence, one may view the connection $A_i$ as the projection operator of the
variation of the ground states onto the Hilbert space orthogonal to them.
It follows from the above definition that $A_i$ is a metric connection with
respect to the Hermitian inner product $g_{i\ol j}$:
\eqn\eMETC{\eqalign{D_ig_{j\ol k} &= \l(\p_i\<\ol k|\r)|j\> + \<\ol
k|\l(\p_i|j\>\r)
- A_{ij}{}^lg_{l\ol k} - g_{j\ol l}A_i{}^{\ol l}{}_{\ol k}\cr
&= 0,\cr}}
where $D_i$ denotes the covariant derivative satisfying
\eqn\eCOVDER{\<\ol k|D_i|j\> = \<\ol k|\l(\p_i - A_i\r)|j\> = 0,}
and
\eqn\eMETCi{A_{ij}{}^k = A_{ij\ol l}g^{\ol lk},\qquad A_i{}^{\ol j}{}_{\ol k}
= g^{\ol jl}A_{il\ol k}.}
Here $g^{\ol ij}$ are the components of $g^{-1}$.

Just as we have defined a connection and covariant derivative in terms of the
variation of the supersymmetric ground states with respect to the holomorphic
parameters in the superpotential $W\l(X_j\r)$, we may likewise define an anti-
holomorphic connection $A_{\ol i}$ and covariant derivative $\ol D_i$,
under
which the metric $g_{i\ol j}$ is also covariantly constant.

Having defined these covariant derivatives, we can compute the curvature of
this connection.
The result of this computation is:
\eqn\eCURV{\eqalign{\Bigl[D_i,D_j\Bigr] &= \l[\ol D_i,\ol D_j\r] =
0,\cr
\l[D_i,\ol D_j\r] &= -\l[C_i,\ol C_j\r],\cr}}
where we define
\eqn\eCBAR{\ol C_j = g\l(C_j\r)^\dagger g^{-1}.}
One also finds that
\eqn\eDC{\eqalign{D_iC_j = &D_jC_i,\qquad\ol D_i\ol C_j = \ol D_j\ol C_i,\cr
&D_i\ol C_j = \ol D_iC_j = 0.\cr}}
These relations have been derived both from a direct analysis of the
supersymmetric ground states of Landau-Ginsburg models \rCGPii, and from
straightforward path integral arguments which apply to any $N=2$
supersymmetric model \rCV.
Hence we see that the curvature is determined by the operator product
coefficients, $\l(C_i\r)_j{}^k$, of the chiral ring ${\cal R}$, as defined in
\eEQV.
By choosing a holomorphic basis for the chiral ring we can set
\eqn\eCHB{A_{i\ol j}{}^{\ol k} = A_{\ol ij}{}^k = 0,}
and this allows us to easily solve for the connection in terms of the metric
$g_{i\ol j}$:
\eqn\eSFM{A_{ij}{}^k=-g_{j\ol l}\l(\p_ig^{-1}\r)^{\ol lk}.}
We then substitute this expression for the connection into the equation for
the curvature, \eCURV, to arrive at a differential equation for the metric,
\eqn\eMASEQ{\ol\p_{\ol i}\l(g\p_ig^{-1}\r) = \l[C_j,g\l(C_i\r)^\dagger
g^{-1}\r],}
which is valid in any holomorphic basis.
{}From \eDC\ we find that the operator product expansion coefficients satisfy
\eqn\DCi{\p_iC_j-\p_jC_i+\l[g\l(\p_ig^{-1}\r), C_j\r]-\l[g\l(\p_jg^{-1}\r),
C_i\r]=0,}
with a similar relation for the derivatives of the $\ol C_i$ matrices.
This equation allows us to nonperturbatively calculate the metric $g_{i\ol
j}$.
If we know its initial value at some point in the space of superpotential
parameters, as well as the first derivatives of the metric at that point, then
in principle we may calculate the metric everywhere by solving the above
second-order partial differential equations.

\newsec{Renormalization Group Flow and the Ramond Charge Matrix}

We now turn to the question of how the $N=2$ theory behaves under the action
of the renormalization group.
The renormalization group flow is a set of trajectories in the space of
quantum field theories relating different quantum field theories via scale
transformations.
The parameter of these curves can be thought of as the ultra-violet cut-off of
the theory, or some other mass scale used to define the quantum field theory.
The action of the renormalization group on the correlation functions of the
theory may be described by saying that the effect of a scale transformation on
the theory is equivalent to an appropriate redefinition of the fields and the
couplings in the theory, determined by the anomalous dimensions and
$\beta$-functions.
An important tool for understanding renormalization group flow in $N=2$
Landau-Ginsburg models is the non-renormalization theorem.
The statement of the non-renormalization theorem is that the only kind of
renormalization that occurs in the superpotential is wave function
renormalization.
While this theorem has only been proven to all orders in perturbation theory,
we shall assume it holds even nonperturbatively.

As an example of how this non-renormalization theorem works, consider the
superpotential with a single chiral superfield $Y$:
\eqn\eWY{\int d\/^2z\,d\/^2\th\ W\l(Y\r) = \int d\/^2z\,d\/^2\th \l({1\ov
4}Y^4 - \ka Y^3\r).}
First consider the case with $\ka=0$.
Under a scale transformation we have $z\rightarrow\la z$,
$\th\rightarrow\la^{-1/2}\th$, so that the term involving the
superpotential
scales as
\eqn\eSPS{\int d\/^2z\,d\/^2\th\ W\l(Y\r) \rightarrow \la\int
d\/^2z\,d\/^2\th\ W\l(Y\r).}
We can eliminate the effect of this scale transformation by making the field
redefinition $Y\rightarrow\la^{-1/4}Y$.
This redefinition will of course change the kinetic term, but we see that the
superpotential is invariant under a combined scale transformation and field
renormalization.
{}From this scaling argument we see that the field $Y$ has an anomalous
dimension of one-fourth, and at the critical point it has conformal weights
$\l(h,\ol h\r)=\l({1\ov8},{1\ov8}\r)$.
This invariance argument holds for any quasi-homogeneous superpotential.

Now consider turning on the coupling $\ka$ and treating it as a perturbation
of the above superpotential.
Under the above scale transformation and field renormalization we find that
\eqn\eRNSP{W\l(Y\r)\rightarrow {1\ov4}Y^4 - \tilde\ka Y^3,}
where $\tilde\ka$ is the renormalized coupling,
\eqn\eKAP{\tilde\ka = \la^{1/4}\ka.}
Hence we see that the $\beta$-function of $\ka$ is determined by the
renormalization of the field $Y^3$.
Once $\ka$ is nonzero, the scaling dimensions of the fields will depend upon
$|\ka|$, but it is intuitively clear that in the large $\la$ limit $\tilde\ka$
will grow and the $Y^3$ term will dominate the superpotential.

We can generalize this picture and take as the action of the renormalization
group the rescaling of the superpotential by a factor $\la$.
The limit $\la\rightarrow 0$ is the ultraviolet limit, and the
$\la\rightarrow\infty$ limit corresponds to the infrared regime.
Since the correlation functions in the topological sector only depend on the
superpotential, we may then use this scaling argument to calculate their
dependence upon the renormalization group parameter.
In the case of Landau-Ginsburg theories, the operator product coefficients
$C_{ij}{}^k$ clearly do not depend on the scaling parameter $\la$, since the
equations of motion \eEOMC\ are independent of $\la$.
Therefore all the dependence on the renormalization group parameter is encoded
in the metric $g_{i\ol j}$.

In the above discussion of the scale dependence of the superpotential, the
parameter $\la$ that appears in front of the superpotential can be taken to be
complex.
However, the metric $g_{i\ol j}$ is independent of the phase of $\la$ since we
can absorb the phase of $\la$ by redefinition of the $\th$ superspace
coordinates in the measure.
This of course leads to a redefinition of the fermionic components of the
chiral superfield $X$, but the inner product $g_{i\ol j}$ depends only on the
lowest bosonic component of the superfield, hence it remains invariant under
such a redefinition.

Consideration of how $g_{i\ol j}$ depends on $|\la|$ near the critical point
leads to the definition of an interesting quantity, the Ramond charge matrix.
{}From \eGIJ\ we have
\eqn\eGL{g_{i\ol j}=\int \prod_{l=0}^n dX_l\,d\ol X_{\ol l}\
\ph_i\l(X\r)\,\ol\ph_{\ol k}\l(\ol X\r)\,\exp\l(\la W-\ol\la\ol W\r).}
At a critical point, the superpotential is quasi-homogeneous, and we find the
dependence of the metric on $|\lambda|$ by a change of variables
$X_i\rightarrow
\la^{q_i}X_i$, giving
\eqn\eGii{g_{i\ol i} \propto \l(\la\ol\la\r)^{-\l(q_i-{\hat c\ov2}\r) -
{n\ov2}},}
where $q_i$ denotes the Neveu-Schwarz $U\l(1\r)$ charge of the superfield
$\ph_i(X)$, $n$ is the number of chiral superfields, and $\hat c= c/3$ is one-
third the value of the conformal anomaly of the theory.
In the above expression we have also used the relation:
\eqn\eCHAT{\hat c = \sum_l\l(1-2q_l\r).}
We note that the quantity $q_i-{\hat c\ov2}$ is the $U\l(1\r)$ charge of the
Ramond ground state which is the image of the Neveu-Schwarz state created by
$\ph_i(X)$ under the action of spectral flow.

We are thus led to define the Ramond charge matrix as
\eqn\eRCM{q_i{}^j = g_{i\ol k}\p_\tau g^{\ol kj} - {n\ov2}.}
The Ramond charge matrix $q$ has a simple field-theoretical interpretation in
terms of the expectation value of a partially conserved charge between Ramond
ground states \refs{\rCGPii,\rCV}.
For simplicity consider a superpotential involving one chiral superfield:
\eqn\eEXSUP{W\l(X\r)= \sum_{i=0}^m t_iX^i,}
and the $R$-symmetry $X\l(z,\th\r)\rightarrow
\e^{i\phi/m}X\l(z,\e^{-i\phi/2}\th\r)$, where $m$ is the largest power
appearing in the superpotential.
When the superpotential is homogeneous, so that $t_i=0$ for $i<m$, then the
current corresponding to this symmetry, $R_{\mu}$, is conserved, otherwise it
is partially conserved.
The Ramond charge matrix is then given by the matrix elements of this
partially conserved charge:
\eqn\eRCHG{q_{k\ol j}=\<\ol j|\oint {d\sigma\ov 2\pi i}R_0\l(\sigma\r)|k\>.}
It can be shown that these matrix elements do not depend on the explicit form
of the superpotential \refs{\rCGPii,\rCV}, (the dependence on $m$ of the
integrated current is in fact $Q^+$-exact), and at the ultraviolet critical
point this matrix is diagonal, the eigenvalues being the Ramond charges of
the supersymmetric ground states, ranging from $-\hat c/2$ to $\hat c/2$.
Therefore at the critical points the maximum eigenvalue of this matrix is
one-sixth the value of the conformal anomaly.
It can also be shown that at the conformal points, where the superpotential is
quasi-homogeneous, $q$ is critical as a function of the coupling constants,
and conversely, the criticality of $q$ implies that the superpotential is
quasi-homogeneous.

These properties of the Ramond charge matrix have led to the speculation that
the maximum eigenvalue of this matrix may provide a ``$c$-function'' on the
space of $N=2$ supersymmetric theories, similar to that defined by
Zamolodchikov on the space of two-dimensional quantum field theories using the
correlation functions of the energy-momentum tensor \ref\rZi{A.B.
Zamolodchikov, JETP Lett. 43 (1986) 730.}.
Both agree at the critical point and behave similarly in a neighborhood of the
critical point.
However we do not have a general argument that the maximum eigenvalue of $q$
is a non-increasing function of the renormalization group flow parameter,
although this has been the case in all models studied so far.
Also the function defined by $q$ is independent of the kinetic $D$-terms in
the
Lagrangian, since it is computed entirely in the topological sector, while the
$c$-function defined by Zamolodchikov does depend on the details of the
$D$-terms.
Hence one might expect that these two definitions might agree for some
particular choice of a $D$-term, but the  precise relationship between these
two
natural functions on the space of $N=2$ quantum field theories is still
unknown.

Another interesting question is to determine the precise relationship between
the metric on the space of two-dimensional quantum field theories defined by
Zamolodchikov in terms of the two-point functions of the perturbing fields
\refs{\rZ,\rZi}, and the inner product of the supersymmetric ground states
$g_{i\ol j}$.
At the critical point the relation between these two metrics is known \rCV.
The inner product discussed above is then related to the two-point function of
the {\it lowest} components of the perturbing superfields evaluated on the
sphere in the Neveu-Schwarz sector:
\eqn\eNSM{{g_{i\ol j}\ov g_{0\ol0}} = G_{i\ol j}=\<\ol\ph_{\ol
j}\l(1\r)\ph_i\>.}
while the metric defined by Zamolodchikov is given by
\eqn\eZAM{G^Z_{i\ol j} = \l\<\int d\/^2\ol\th\ \ol\Phi_{\ol j}\l(1\r)\,\int
d\/^2\th\ \Phi_i\l(0\r)\r\>,}
where $\ph_i$ is the lowest component of the superfield $\Phi_i$.
The factor of $g_{0\ol0}^{-1}$, where the index $0$ labels the identity,
occurs in \eNSM\ to provide the correct normalization by dividing out the
vacuum amplitude.
At the conformal point one may directly relate these two metrics, and using
the superconformal Ward identities one finds
\eqn\eNSMZAM{G^Z_{i\ol j}=q_{i,L}q_{i,R}G_{i\ol j},}
where $q_{i,L}$ and $q_{i,R}$ are the left and right $U(1)$ charges.
In particular, we see that the components of the Zamolodchikov metric
involving the identity vanish, since perturbations by multiples of the
identity are annihilated by the integration over superspace, and for marginal
perturbations, satisfying
$q_{i,L}=q_{i,R}=1$, the metrics are identical.

These considerations lead to the definition of the algebraic $Q$-matrix \rCV,
\eqn\eQMAT{Q_i{}^j = G_{i\ol k}\p_{\tau}\l(G^{-1}\r)^{\ol kj},}
where $G$ is the normalized matrix defined in \eNSM.
At the critical point, the eigenvalues of $Q$ are the $U(1)$ charges of the
chiral primary fields, ranging from $0$ to $\hat c = c/3$.
It was suggested in \rCV\ that the maximum eigenvalue of this matrix might
also
be a candidate $c$-function, but we shall provide an example below in which
the infrared critical limit of the maximum eigenvalue of $Q$ is approached
from below, and hence is not a non-increasing function of the renormalization
group flow parameter.

It is interesting to note that the definition of the $Q$-matrix \eQMAT\ is
quite similar to the renormalization group equation for the two-point function
$\<\ol\phi_{\ol j}\l(1\r)\phi_i\l(0\r)\>$.
Under the scale transformation $x\rightarrow \e^{|\tau|}x$, the two-point
function satisfies the equation
\eqn\eRGEQ{\l(\half{\p\ov\p|\tau|}+\what\Ga -
\beta^a{\p\ov\p t^a}\r)\<\ol\phi_{\ol j}\l(1\r)\phi_i\l(0\r)\>=0,}
where $\what\Ga$ is the anomalous dimension operator
\eqn\eADO{\what\Ga\ph_i = \ga_i{}^j\ph_j,\qquad \what\Ga\ol\ph_{\ol i} =
\ol\ph_{\ol j}\ga^{\ol j}{}_{\ol i},}
and the coefficients $\beta^a$ are the $\beta$-functions related to the scale-
dependent coupling constants by
\eqn\eBETA{\beta^a = \half{dt^a\ov d|\tau|}.}
The anomalous dimension matrix has been normalized such that its eigenvalues
are the conformal weights at the critical point, equal to one-half the scaling
dimension for Landau-Ginsburg models, and the sum over $a$ runs over all the
coupling constants in the theory.

If we now conjecture that the identification in \eNSM\ holds away from the
critical point as well, (perhaps this identification again corresponds to a
specific choice of a kinetic $D$-term), then since the left-hand side of
\eNSM\
depends only on the superpotential, we may restrict the sum over coupling
constants to those in the superpotential alone.
The choice of a holomorphic basis allows us to write
\eqn\eSBET{\eqalign{\beta^a\p_a G_{i\ol j} &=\beta^k\p_kG_{i\ol j}+
\beta^{\ol k}\p_{\ol k}G_{i\ol j}\cr
&=\beta^k{\cal A}_{ki}{}^lG_{l\ol j} +
G_{i\ol l}\beta^{\ol k}{\cal A}_{\ol k}{}^{\ol l}{}_{\ol j},\cr}}
where ${\cal A}_i$ is the metric connection with respect to $G_{i\ol j}$:
\eqn\eCALA{{\cal A}_{ij}{}^k = -G_{j\ol l}\l(\p_iG^{-1}\r)^{\ol lk}.}
If we rewrite the definition of $Q$, \eQMAT, as follows,
\eqn\eQMATi{\p_{\tau}G_{i\ol j} + \half\l(Q_i{}^kG_{k\ol j}+
G_{i\ol k}Q^{\ol k}{}_{\ol j}\r)=0,}
then since
\eqn\eTRIV{\p_{\tau}G_{i\ol j}=\half{\p\ov\p|\tau|}G_{i\ol j},}
we are led to the identification:
\eqn\eQREL{\half Q_i{}^j = \ga_i{}^j - \beta^k{\cal A}_{ki}{}^j.}
Hence the $Q$-matrix seems to be directly related to the anomalous dimensions
of the chiral fields, the $\beta$-functions of the theory, and the field-
theoretical connection with respect to the normalized metric $G_{i\ol j}$.
In particular, at the critical points, where the $\beta$-functions vanish, we
see that one-half the eigenvalues of $Q$ are the conformal weights of the
chiral primary fields, as expected.

\newsec{Renormalization Group Flow Along a Critical Line}

We shall now apply the above formalism to a specific model, characterized by
the superpotential involving one chiral superfield:
\eqn\eSUP{W(X) = {1\ov4}X^4 - \alpha X- \beta X^2,}
where $\alpha$ and $\beta$ are complex parameters.
The chiral fields $X$ and $X^2$ are two of the three relevant superfields in
the theory, the third being the identity, which does not deform the
superpotential.
The cases for $\alpha=0$ or $\beta=0$ have previously been studied by Cecotti
and Vafa \rCV.
These correspond to massive deformations of the $N=2$ $A_3$ minimal model with
$\hat c=\half$.
In both cases, the deformation gives an integrable model, and there exists a
basis of the chiral ring for which the metric $g_{i\ol j}$ is diagonal and can
be parameterized by a single function.
The second order differential equation that results from \eMASEQ\ in these
cases turns out to be special cases of the third Painlev\'{e} equation .

In this section we shall instead consider a perturbation which retains a
massless field in the infrared limit.
Recall that the bosonic potential $V\l(X\r)$ is given in terms of the
superpotential by:
\eqn\eBOSP{V\l(X\r) = \bigg|{\p\ov\p X}W\l(X\r)\bigg|^2.}
The classical ground states correspond to the critical points of $W$, where
the partial derivative vanishes.
A requirement of the superpotential for a massless field to exist is that at
one of the critical points the Hessian, defined by \eHESS, also vanishes, so
that at least two critical points will be degenerate.
Imposing this constraint on the above superpotential \eSUP, and solving the
two equations $\p W = H = 0$ gives the following relation between $\alpha$ and
$\beta$:
\eqn\eCONS{\l({1\ov2}\alpha\r)^2 = \l({2\ov3}\beta\r)^3 \equiv \kappa^6.}
If we make the field redefinition $\tilde Y=X+\kappa$, then aside from an
irrelevant constant the superpotential takes the form
\eqn\eSUPii{\tilde W\l(\tilde Y\r) = {1\ov4}\tilde Y^4 - \kappa \tilde Y^3.}
This superpotential has one critical point at $\tilde Y=3\kappa$ and two
which lie at $\tilde Y=0$.
In accord with our previous discussion of the renormalization group flow of
this model, we expect the renormalized coupling $\kappa$ to scale as
$\lambda^{1/4}$, and therefore in the limit $\lambda\rightarrow\infty$ the two
sets of ground states should decouple, with the double degeneracy at $\tilde
Y=0$ being described by the $N=2$ $A_2$ minimal model, with exponentially
suppressed overlap with the massive theory at $\tilde Y=3\kappa$.

We wish to calculate the metric $g_{i\ol j}$ as a function of the
renormalization group scale $\lambda$, so it is convenient to perform one more
field redefinition so that the dependence on $\lambda$ is explicit.
Let $\kappa = \la^{1/4}$ and $Y = \la^{-{1/4}}\tilde Y$ so that the
superpotential now takes the form
\eqn\eYPOT{W\l(Y\r) = \la\l({1\ov4}Y^4 - Y^3\r).}
Our strategy for calculating the metric $g_{i\ol j}$ is as follows.
First we shall solve the constraint \eRCON\ and require that the metric be
Hermitian by finding a convenient parameterization of the metric.
We shall then calculate the metric as a function of the renormalization group
flow parameter $\la$ by solving the differential equation \eMASEQ\ using the
appropriate initial conditions on the metric at $\la=0$.
We may then calculate the correlation functions in the topological sector as
functions of the scale parameter $\lambda$, and verify our expectations based
on the $N=2$ non-renormalization theorem concerning the infrared behavior of
this model.

For our purposes, a convenient basis is given by
\eqn\eBASIS{\l\{\ph_i\r\} = \l\{1,\l(Y -1\r),
\l(Y^2 - 2Y -\half\r)\r\},\qquad i=0,1,2.}
We shall call this basis the flat basis.
In this basis the topological inner product has a very simple dependence on
the parameters in the superpotential, and is simply given by
\eqn\eNMAT{\eta = \la^{-2}\pmatrix{0&0&1\cr
                           0&1&0\cr
                           1&0&0\cr}.}
Using \eEQV\ we may calculate the matrices $C_i$:
\eqn\eCMAT{C_0 = \pmatrix{1&0&0\cr 0&1&0\cr 0&0&1\cr},\
C_1 = \pmatrix{0&1&0\cr 3/2&0&1\cr 2&3/2&0\cr},\
C_2 = \pmatrix{0&0&1\cr 2&3/2&0\cr 9/4&2&0\cr},}
which appear in the topological correlation functions as in \eCGF.

The key to solving the reality constraint is to notice that in the flat basis
the above topological inner product \eNMAT\ implies that \eRCON\ takes the
form
\eqn\eRCONi{ g\eta g^T = \eta,}
so the metric $g_{i\ol j}$ is essentially an element of complexified
$SO(2,1)$, and therefore may be parameterized by three complex Euler angles.
The parameterization used in the calculation of the metric is given by
\eqn\eGii{g=ST\tilde gT^\dagger S^\dagger,}
where
\eqn\eGi{\tilde g = \pmatrix{\cos\psi&\sin\psi&0\cr
-\sin\psi&\cos\psi&0\cr
0&0&1\cr}
\pmatrix{1&0&0\cr
0&\cosh\th&\sinh\th\cr
0&\sinh\th&\cosh\th\cr}
\pmatrix{\cos\rh&\sin\rh&0\cr
-\sin\rh&\cos\rh&0\cr
0&0&1\cr},}
where $\psi$, $\th$, and $\rh$ are three complex parameters.
The matrices $T$ and $S$ are given by
\eqn\eT{T =  T^{-1} = \pmatrix{1/{\sqrt2}&0&1/{\sqrt2}\cr
                                       0&1&0\cr
                               1/{\sqrt2}&0&-1/{\sqrt2}\cr},}
and
\eqn\eDEFS{S = \pmatrix{\la^{-1/4}&0&0\cr
                        0&\la^{-1/2}&0\cr
                        0&0&\la^{-3/4}\cr}.}

Now requiring that $g$ be a Hermitian matrix requires that $\psi=-
\rh$, and that $\th$ be either real or purely imaginary.
We shall see below that the initial conditions on the metric require that
$\th$ be real.
Hence we have a parameterization of the metric in terms of three real
functions of $\la$.

For the superpotential we are considering, \eYPOT, there is actually one
remaining symmetry of the metric $g_{i\ol j}$.
It is invariant under the interchange of $\ph_i$ with $\ol\ph_{\ol i}$, since
the only complex parameter in the superpotential is $\la$, and as discussed in
Section 4, the metric does not depend on the phase of $\la$.
Hence, in the flat basis above, the metric is not only Hermitian, but also
symmetric.
This symmetry implies that $\psi$ must be real, thus in our calculation the
metric may be parameterized by just two real functions of $\la$.

We now wish to solve the for the dependence of the metric on the
renormalization group scale $\la$ by solving the differential equation
\eqn\eDIFFLA{{\p\ov\p\ol\la}\l(g{\p\ov\p\la}g^{-1}\r) = \l[C_{\la},
g\l(C_{\la}\r)^{\dag} g^{-1}\r],}
where the matrix $C_{\la}$ is the matrix representation of the superpotential
itself:
\eqn\eCLA{C_{\la} = {3\ov4}\l(C_2+2C_1+{5\ov2}C_0\r),}
since
\eqn\eCLAi{\eqalign{-{\p W\ov\p\la} &= -\l({1\ov4}Y^4 - Y^3\r) \sim
{3\ov4}Y^2\cr
&\sim {3\ov4}\l[\l(Y^2-2Y-\half\r)+2\l(Y-1\r)+{5\ov2}\r].\cr}}
Since the component of $C_{\la}$ proportional to the identity matrix vanishes
in the commutation relation in the above differential equation, we may neglect
it and simply evaluate
\eqn\eCLAii{C_{\la} = {3\ov4}\pmatrix{0&2&1\cr 5&3/2&2\cr 25/4&5&0\cr}.}
Using the fact that in the flat basis above the metric is only a function of
$x=|\la|$ we finally arrive at the following second order differential
equation for the metric:
\eqn\eFINEQ{{1\ov4x}{d\ov dx}\l(xg{d\ov dx}g^{-1}\r) = \l[C_{\la},gC_{\la}^T
g^{-1}\r].}

To solve this differential equation, we first must specify the initial
conditions on the metric.
Before determining the boundary conditions, it is instructive to first recall
some aspects of the analysis of the two cases where $\alpha=0$ or $\beta=0$
described in \rCV.
In both these cases the differential equation which describes the dependence
of the metric on the parameters $\alpha$ or $\beta$ is the third
Painlev\'{e} transcendent equation:
\eqn\ePTE{f'' = {\l(f'\r)^2\ov f}-{f'\ov x}+{1\ov x}\l(p_1f^2+p_2\r)+
p_3f^3+{p_4\ov f},}
where the function $f$ is essentially given by one of the (diagonal) elements
of the metric and $x$ is related to the perturbing coupling $\alpha$ or
$\beta$.
In the case $\alpha=0$ the parameters $p_i$ take the values $p_1=-p_4=1$,
$p_2=p_3=0$, and in the case where $\beta=0$ we have $p_1=p_2=0$, $p_3=-
p_4=1$.
This equation has been studied using the isomonodromic deformation method in
\nref\rIN{A.R. Its and V.Yu. Novokshenov, {\it The Isomonodromic
Deformation Method in the Theory of Painlev\'{e} Equations}, Lecture Notes in
Mathematics 1191, Springer-Verlag, Berlin 1986.}
\nref\rKv{A.V. Kitaev, {\it The Method of Isomonodromy Deformations for
Degenerate Third Painlev\'{e} Equation}, published in {\it Questions of
Quantum Field Theory and Statistical Physics 8} (Zap. Nauch. Semin. LOMI
161), P.P. Kulish and V.N. Popov (ed.), Nauka, Leningrad.}
\refs{\rIN,\rKv}.
Using the results of these studies, S. Cecotti and C. Vafa found that the
requirement that $g_{i\ol j}$ be a nonsingular, positive-definite metric will
uniquely determine the required boundary conditions of the solution to the
above differential equation.
This remarkable situation was also found to be the case in many other models
they studied, all of which correspond to integrable deformations of the $N=2$
minimal models and are related to quantum and classical affine Toda theories.
Hence the physical requirement of the regularity of the solution, combined
with the differential equations \eMASEQ, may completely determine the metric,
including its values at the critical point, which gives the normalized values
of the operator product expansion coefficients of the corresponding conformal
field theory.

Since in the flat basis above the metric $g_{i\ol j}$ is singular as
$\la\rightarrow 0$, we shall discuss the initial conditions in the basis
$\l\{1, X, X^2\r\}$ for the superpotential in the form \eSUP\ where the
parameters $\alpha$ and $\beta$ satisfy the relation \eCONS.
The differential equations for the functions $\psi$ and $\th$ which result
from \eFINEQ\ were determined using a mathematical manipulation language,
Maple.
The differential equations which result are rather lengthy, and do not appear
to have previously been studied in the literature, so we have not yet
determined whether the requirement of regularity alone uniquely determines the
initial data for the metric solution.
However, using the known value of the metric for the unperturbed conformal
theory, we have been able to determine the initial values of the first
derivatives of the parameters $\psi$ and $\th$ by requiring that the
differential equations for these parameters are nonsingular at the origin
$x=|\la|=0$.

Using some results derived in \rSCi, the relevant ones having been collected
in Appendix A below, we find
\eqn\eICMET{g^X_{i\ol j}\l(x=0\r) = \pmatrix{\ga/2&0&0\cr 0&1&0\cr
0&0&2/\ga\cr},}
where $x=|\la|$ and $\ga$ is the ratio of the gamma functions:
\eqn\eICi{\ga = {\Ga\l(1/4\r)\ov\Ga\l(3/4\r)}.}
The initial conditions on the derivatives of the metric elements at
$x=|\la|=0$ are found to be:
\eqna\eICDER
$$\eqalignno{{d\ov dx}\<\ol0|0\> &=-{9\ov4}\<\ol2|2\>,&\eICDER a\cr
{d\ov dx}\<\ol2|2\> &={9\ov4}\l[\l(\<\ol2|2\>\r)^3 + \<\ol0|0\>\r],&\eICDER
b\cr
{d\ov dx}\<\ol0|2\> &={3\ov4}x^{-1/2}\e^{i2\phi}\<\ol0|0\>,&\eICDER c\cr
{d\ov dx}\<\ol2|0\> &={3\ov4}x^{-1/2}\e^{-i2\phi}\<\ol0|0\>,&\eICDER d\cr}$$
where the phase $\phi$ arises due to the trivial dependence of the metric
elements on the phase of $\kappa=|\kappa|\e^{i\phi}$ in the above basis.
The metric elements referred to on the right-hand side of \eICDER\ are those
of the metric at $x=0$, \eICMET.
The first derivatives of all other elements vanish at $x=0$.
These initial conditions on the first derivatives of the metric agree with
those found by Cecotti and Vafa in \rCV.

\newsec{Numerical Solution and Comparison with the Asymptotic Solution}

Applying the initial data on the metric given above, we have solved the
differential equation \eFINEQ\ numerically, using the fourth-fifth order
Runge-Kutta method implemented by Maple.

In \fig\fqr{Eigenvalues of the Ramond charge matrix as a function of the
coupling $\kappa$} we have plotted the eigenvalues of the Ramond charge
matrix, calculated in the flat basis \eBASIS, as a function of the perturbing
coupling $\kappa$ in the superpotential.
Recall that the running coupling $\kappa$ is related to the renormalization
group scale parameter $\la$ by $\kappa=\la^{1/4}$.
At $\kappa=0$ the maximum eigenvalue is one-fourth, equal to the value of
$\hat c/2$ of the unperturbed theory.
In the infrared limit we see that the maximum eigenvalue monotonically
decreases to the value one-sixth, the value of $\hat c/2$ for the theory
described by the superpotential $W\l(Y\r) = Y^3$.
In this limit, the two eigenvalues approaching $\pm{1\ov6}$ correspond to the
Ramond charges of the two supersymmetric ground states of the conformal
theory, and the zero eigenvalue corresponds to the trivial massive theory
which is decoupled.
We observe from this graph that we expect the infrared limit of the theory to
behave critically near the value $|\kappa|=0.6$.

In \fig\fQM{Eigenvalues of the algebraic $Q$-matrix as a function of the
coupling $\kappa$} we have plotted the eigenvalues of the $Q$-matrix, again
calculated in the flat basis, as a function of $\kappa$.
These eigenvalues exhibit a different behavior than those of the Ramond charge
matrix.
Using the asymptotic solution for the metric, discussed in more detail below,
we may compute the infrared limit behavior of the $Q$-matrix and its
eigenvalues.
In the flat basis we find that as $|\kappa|\rightarrow\infty$:
\eqn\eQAS{Q\approx{1\ov36}\pmatrix{{\l(12|\kappa|^{4/3}\chi-
10\r)\ov\l(6|\kappa|^{4/3}\chi+1\r)}&2&0\cr
5&{\l(36|\kappa|^{4/3}\chi-6\r)\ov\l(6|\kappa|^{4/3}\chi+1\r)}&-2\cr
0&-5&{\l(60|\kappa|^{4/3}\chi-2\r)\ov\l(6|\kappa|^{4/3}\chi+1\r)}\cr},}
where
\eqn\eDEFCHI{\chi={\Ga\l(1/3\r)\ov\Ga\l(2/3\r)}.}
The eigenvalues of this matrix approach the asymptotic values
$\l(0,1/6,1/3\r)$ from below, and for $|\kappa|>0.74$ are in good agreement
with the calculated values.
The eigenvalues $\l(0,1/3\r)$ correspond to the $U(1)$ charges of the chiral
fields in the infrared conformal theory, while the eigenvalue of one-sixth
corresponds
to the massive theory, one-sixth being the shift due to spectral flow.
It is interesting to note that the asymptotic form of the algebraic $Q$-matrix
depends on explicitly on $\kappa$, while the similar asymptotic form of the
Ramond charge matrix, given by
\eqn\eQRAS{q_R\approx{1\ov36}\pmatrix{-4&2&0\cr 5&0&-2\cr 0&-5&4\cr},}
in the flat basis, is independent of $\kappa$.

In order to compare the results of the computation of the components of the
metric with the asymptotic
solution, we have chosen to display the results of the calculation of the
metric elements in the ``point'' basis.
Physically, this is the basis in which each element of the basis corresponds
to a ground state wave function localized near a critical point of the
superpotential.
When the superpotential is of the form $\tilde W(\tilde Y)$, \eSUPii, the
point basis is given by
\eqn\ePBAS{\l\{\ph_i^P\r\}=\l\{{1\ov3\kappa}\l(3\kappa-\Yt\r),
{1\ov3\kappa}\Yt\l(3\kappa-\Yt\r), {1\ov9\kappa^2}\Yt^2\r\},\qquad i=0,1,2.}
In this basis the matrices $C_i$ and the topological inner product $\eta_{ij}$
are in block diagonal form, the first two elements of this basis representing
the doubly degenerate critical point at $\Yt=0$ and the third element
representing the isolated critical point at $\Yt=3\kappa$.
Asymptotically the metric $g_{i\ol j}$ will be in block diagonal form with
exponentially suppressed off-diagonal metric elements.

In the limit $\kappa\rightarrow\infty$ the superpotential behaves as $\tilde
W\rightarrow -\kappa\Yt^3$, and the first two elements of the point basis give
a basis for the chiral ring of the infrared conformal theory:
\eqn\ePBASi{\l\{{1\ov3\kappa}\l(3\kappa-\Yt\r),
{1\ov3\kappa}\Yt\l(3\kappa-\Yt\r)\r\}\rightarrow\l\{1,\Yt\r\}.}
Again using the results of Appendix A, we may use the above behavior of the
first two elements of the point basis to determine the leading dependence on
$\kappa$ of the metric components $\<\ol0|0\>_P$ and $\<\ol1|1\>_P$ in the
point
basis.
Then using the reality constraint \eRCON, where
\eqn\eNPT{\tilde\eta_P={-1\ov3\kappa}\pmatrix{-1/3\kappa&1&0\cr 1&0&0\cr
0&0&-1/3\kappa\cr},}
in the point basis, we find that the asymptotic limit of the metric in this
basis is
\eqn\eAMETPT{\tilde g_P=\pmatrix{{1\ov3}\l(|\kappa|^{-2/3}\chi+
|\kappa|^{-10/3}{1\ov36\chi}\r)&-|\kappa|^{-7/3}\e^{-i\phi}{1\ov18\chi}&0\cr
-|\kappa|^{-7/3}\e^{i\phi}{1\ov18\chi}&|\kappa|^{-4/3}{1\ov3\chi}&0\cr
0&0&|\kappa|^{-2}{1\ov9}},}
where $\chi$ was defined above in \eDEFCHI, and
$\kappa=|\kappa|\e^{i\phi}$ is the coupling in the superpotential, and we have
neglected exponentially suppressed terms.

In \nfig\fGoo{Plot of $g_{0\ol0}$ (in the point basis) as a function of the
coupling $\kappa$}\nfig\fGoi{Plot of $|g_{0\ol1}|$ (in the point basis) as a
function of the coupling $\kappa$}\nfig\fGii{Plot of $g_{1\ol1}$ (in the point
basis) as a function of the coupling $\kappa$}\nfig\fGtt{Plot of $g_{2\ol2}$
(in the point basis) as a function of the coupling $\kappa$}\figs{\fGoo{--
}\fGtt} we have plotted the calculated values of the metric components in the
point basis as functions of the coupling $\kappa$, as shown by the solid
curves.
The dashed curves shown are the asymptotic values of the metric, given in
\eAMETPT\ above, and in every case we see excellent agreement for
$|\kappa|>0.6$.

We can also compare the off-diagonal metric components, $\<\ol2|0\>$ and
$\<\ol2|1\>$, to the leading order semi-classical corrections to \eAMETPT.
These components of the metric may be viewed as arising from the probability
to tunnel between the
classical ground states localized near distinct critical points.
It has been argued in \rCV\ that in a basis in which the metric is diagonal to
leading order, the leading off-diagonal semi-classical correction to the
metric is a universal function of the mass of the soliton connecting the two
distinct critical points, if it exists.
In such a basis, this function is of the form
\eqn\eSCC{{g_{i\ol j}\ov(g_{i\ol i}g_{j\ol j})^{1/2}}\approx C_{i\ol
j}m_{ij}^{-1/2}\exp\l(-m_{ij}\r),}
where $C_{i\ol j}$ is some complex coefficient and $m_{ij}$ is the mass of the
soliton interpolating between the two vacua:
\eqn\eSOLM{m_{ij}=2|\lambda W\l(X_i\r)-\lambda W\l(X_j\r)|,}
$X_i$ and $X_j$ being the values of the distinct critical points.
The method of calculating the above semi-classical correction is discussed in
Appendix B.
In the model we are analyzing the mass of the soliton connecting the vacua at
$\tilde Y=0$ with $\tilde Y=3\kappa$ is $m=27|\kappa|^4/2$, and it is of order
one for
$|\kappa|\approx0.52$, a result in excellent accord with the behavior of the
metric components we have displayed.

\nfig\fGot{Plot of $-\ln\l(|g_{0\ol2}|\r)$ (in the point basis) as a function
of the coupling $\kappa$}
\nfig\fGit{Plot of $-\ln\l(|g_{1\ol2}|\r)$ (in the point basis) as a function
of the coupling $\kappa$}
We have plotted the logarithm of the final two components of the metric in
\figs{\fGot,\fGit}, and
have displayed the leading semi-classical correction by a dashed line.
These semi-classical corrections are given in the point basis by:
\eqn\eGoi{|g_{0\ol2}|={(0.76)\ov162(\pi\chi)^{1/2}}|\kappa|^{-
14/3}\l(6\chi|\kappa|^{4/3}-1\r)F(m)\exp(-m),}
\eqn\eGoii{|g_{1\ol2}|={(0.76)\ov27(\pi\chi)^{1/2}}|\kappa|^{-11/3}F(m)\exp(-
m),}
where $m$ is the mass of the soliton defined above, and $F(m)$ is a function
defined in Appendix B.
Unfortunately the method of calculating the form of these corrections does not
give us the value of the constant factor $C_{i\ol j}$, so the number $0.76$ in
\eGoi\ and \eGoii\ was determined by a fit to the calculation.
In principle these corrections might also be calculated by using the WKB
approximation to compute the overlap of the wave-functions localized near
distinct critical points, but unfortunately this method also becomes
unreliable in the neighborhood of the critical points of the superpotential
and cannot be used to determine this factor.
Nevertheless, after fitting this one parameter we again see good agreement
between calculation and the predicted value of the soliton mass.

There are two primary sources of error in the numerical computations we have
presented.
One is associated with the finite step width involved in the numerical
solution of the differential equations.
The relative and absolute error tolerances in the computed quantities were
chosen to be no greater than $10^{-11}$.
However, by varying the allowed tolerances over several orders of magnitude,
it was found that the numerical results were quite insensitive to the step
width involved in the range over which calculations were performed.

The other source of error is associated with the initial conditions on the
metric and its first derivatives.
In the parameterization of the metric used in the computations, while the
initial values and the first derivatives of the parameters used are finite at
the point $|\lambda_0|=0$, the second derivatives of the parameters are
singular at this initial point.
For this reason, we were required to specify the initial data at some point
off the origin, and the initial point was taken to be $|\lambda_1|=10^{-6}$.
Unfortunately, the precise value of the metric and its derivatives is not
known away from the origin, so the initial data was corrected to first order
in $|\lambda_1-\lambda_0|=|\lambda_1|$ for the first derivatives and to second
order in $|\lambda_1|$ for the initial values by solving the differential
equations in the small $|\lambda|$ limit.

By studying the effect of variations on the initial data it was found that
both the values and the general behavior of the calculated quantities are very
sensitive to the precise initial conditions imposed.
In fact, it appears that the major source of error is attributable to the
error in the initial conditions.
This is in accord with our general expectations based on the previously
studied cases discussed above.
Unless the boundary conditions agree with the requirement of regularity, then
we would expect to find some singularities in the solution of our differential
equations, which would drastically alter the its nature.
The departure of the calculated values of the off-diagonal metric elements
from the asymptotic limit in \figs{\fGot,\fGit}, for $|\kappa|>0.8$ is the
first indication of this error.
By making minor adjustments in the initial data it is possible to correct this
departure from the asymptotic limit, however we have chosen not to do this,
and instead accept the value $|\kappa|\approx0.8$ as the limit of the
reliability of these calculations.

\newsec{Conclusions}

We have demonstrated by explicit computation using nonperturbative methods
that the correlation functions in the topological sector of a theory with
$N=2$ supersymmetry may be computed both on and off of the critical point.
We studied a particular perturbation of the $A_3$ superconformal minimal model
which in the infrared limit sends the theory to the $A_2$ model along the
renormalization group trajectory.
Using standard numerical techniques we essentially obtained all the
characteristic features of the infrared conformal field theory, such as the
value of the conformal anomaly and the normalized operator product expansion
coefficients of the chiral ring.

We also calculated the amplitude for quantum ``tunnelling'' between the
conformal theory described by the $A_2$ model and the massive theory which
decouples in the infrared limit.
These results were compared to the leading order semi-classical instanton
corrections, and we found good agreement with the predicted value of the mass
of the soliton which connects the two critical points.
Hence we have exhibited the complete behavior of the topological sector of the
theory which interpolates between two nontrivial conformal field theories
along the renormalization group flow trajectory.
We are currently in the process of extending these results to the full space
of perturbations of the $A_3$ model \ref\rLHi{W.A. Leaf-Herrmann, {\it
Nonperturbative Analysis of the Space of Perturbations of an $N=2$
Superconformal Field Theory}, in preparation.}.

Several interesting questions remain.
The physical requirement of the regularity of the Hermitian metric seems to
essentially fix the initial data of the second order differential equation
which describes its dependence on the parameters of the theory.
This was found to be the case in all the models studied in \rCV, and although
it was used to fix only the first derivatives in the problem studied above, it
almost certainly must fix the initial values of the metric as well.
Is this a general feature of all $N=2$ supersymmetric models, and what is the
precise connection with the isomonodromic deformation techniques used by
mathematicians?

It would also be interesting to determine the precise relation between
Zamolodchikov's field theoretical metric and the $c$-function, and the
Hermitian inner product and Ramond charge matrix of the topological sector.
This would lead to a better understanding of the connection between the usual
first order field-theoretical differential equations describing the
renormalization group flow and the second order partial differential equations
used above to describe the behavior of the Hermitian inner product of the
Ramond ground states.

\bigbreak\bigskip\bigskip\centerline{{\bf Acknowledgements}}\nobreak

I would like to thank C. Vafa for bringing this problem to my attention, as
well as for many valuable discussions and comments.
I would also like to acknowledge a helpful discussion with S. Cecotti.
This work was supported in part by NSF grant PHY-87-14654.

\appendix{A}{Metric Components of the Conformal Theory}

In this appendix we collect the results derived in \rSCi\ which were used to
calculate the initial values of the metric components, as well as their
asymptotic limits, in the case when the superpotential is homogeneous.
Consider the case when the superpotential is of the form $W(X)=\lambda X^m$.
Using the fact that field theoretical computations in the topological sector
may be reduced by dimensional reduction to computations in supersymmetric
quantum mechanics, we may express the computation of the metric $g_{i\ol j}$
as the inner product between vacuum waveforms:
\eqn\eIPM{g_{i\ol j} = \<\ol j|i\> = \int *\ol\omega_j\wedge\omega_i,}
where $\omega_i$ is the differential form associated with the supersymmetric
ground state wave function $|i\>$ in the standard fashion \ref\rWD{E. Witten,
J. Diff. Geom. 17 (1982) 661.}.
This representation of $g_{i\ol j}$ can be shown to be equivalent to that
given in \eGIJ\ for a quasi-homogeneous superpotential \rCV.
Choose an orthonormal basis for the ground state waveforms:
\eqn\eBGSW{\omega_i = \alpha_i\lambda^{(i+1)/m}X^i\,dX,\qquad i=0,\ldots,m-
2,}
where the normalization coefficients $\alpha_i$ are fixed up to a phase by the
requirement of orthonormality.
This waveform corresponds to the basis
$\{\ph_i\}=\{\alpha_i\lambda^{(i+1)/m}X^i\}$ of the chiral ring ${\cal R}$.

{}From the real structure matrix $M$, defined in \eDEFM, we have the relation
between a vacuum waveform and its complex conjugate:
\eqn\eRSR{\ol\omega_j = M_{\ol j}{}^i\omega_i.}
Letting $\tilde\omega$ denote the vacuum waveform corresponding to $\ol\omega$
via the real structure, we may write
\eqn\eBGSWi{\tilde\omega_j = \beta_j\lambda^{(m-j-1)/m}X^{m-j-2}\,dX,\qquad
j=0,\ldots,m-2,}
where we have essentially used the action of spectral flow to determine the
power of $X$ which appears, and $\beta_j$ is determined by the real structure.

Using the Bochner-Martinelli theorem it can be shown \rSCi\ that
\eqn\eBMT{\eqalign{g_{k\ol j} &= \int *\ol\omega_j\wedge\omega_k\cr
&=Res_W[\tilde\ph_j(X)\ph_k(X)],\cr}}
and therefore the orthonormality condition implies that
\eqn\eONC{\alpha_k\beta_k=m.}
To determine the normalizations $\alpha$ we need one more condition, the
reality condition expressed by S. Cecotti \rSCi.
If we define $W_1$ to be the points defined by $X^m=1$:
\eqn\eWONE{W_1=\{X_j=\exp(2\pi ij/m)\},\qquad j=0,\ldots,m-1,}
then a basis for the relative homology group $H_1({\bf C}, W_1;{\bf Z})$ is
given by the segments $\gamma_j$ connecting the points $X_{j+1}$ and $X_j$.
Using the reality condition of \rSCi,
\eqn\eSCRC{\int_{\gamma_j} \e^{-W}\omega_k = \l(\int_{\gamma_j} \e^{-W}
\tilde\omega_k\r)^*,}
and the fact that
\eqn\eSINGINT{\int_{\gamma_j}\e^{-W}\omega_k = {2i\ov m}\alpha_k\e^{2\pi
ijk/m}\sin\l({k\pi\ov m}\r)\Ga\l({k\ov m}\r),}
we arrive at the condition that
\eqn\eAOB{{\alpha_k\ov\beta_k} = -{\Ga(1-k/m)\ov\Ga(k/m)}.}
Combining this condition with \eONC\ above we find that
\eqn\eNORP{\alpha_k = \sqrt{n\pi}\l[\sqrt{\sin(k\pi/m)}\Ga(k/m)\r]^{-1}.}

For the superpotential
\eqn\eHSUP{W(X) = {1\ov4}X^4,}
the relation between the above orthogonal basis, $|k\>_O$, and that associated
with the polynomial basis $\{1,X,X^2\}$, say $|j\>_X$, is given by
\eqn\eOTP{|j\>_X = U_j{}^k|k\>_O,}
where
\eqn\eDEFU{U = \hbox{diag}\l(4^{1/4}\alpha_0^{-1},
4^{1/2}\alpha_1^{-1},4^{3/4}\alpha_2^{-1}\r).}
Hence in the polynomial basis we have
\eqn\eGPOLY{g^X = Ug^OU^{\dagger},}
where $g^O$ is simply the identity matrix, due to the orthonormality of the
basis $|k\>_O$.
This equation directly gives \eICMET.
Similarly, for the superpotential $W(\tilde Y)=\kappa\tilde Y^3$, the above
relations combined with the reality constraint \eRCON\ lead to the result
given in \eAMETPT.

\appendix{B}{Leading Semi-Classical Corrections to the Metric}

In this appendix we outline the method used to calculate the semi-classical
corrections given in \eGoi\ and \eGoii.
For a comprehensive discussion of semi-classical considerations, see Appendix
B of \rCV.
Beginning with the metric in the point basis \eAMETPT, one first expresses it
in the form
\eqn\eMETB{\tilde g_P = V\exp\l(\zeta\r)V^{\dagger},}
where the matrix $V$ is the holomorphic transformation to an orthonormal
basis, and $\zeta$ represents the exponentially suppressed off-diagonal terms,
so that classically $\zeta=0$.
In the large $\lambda$ limit the calculation of $\zeta$ may be approximated by
the one-instanton contribution, neglecting other contributions which are
exponentially suppressed with respect to this leading order correction.
Hence we may work to first order in $\zeta$.

In this approximation \eFINEQ\ is given by
\eqn\eZAPP{{1\ov4x}{d\ov dx}\l(x{d\ov dx}\zeta\r) =
\l[C_{\lambda},\l[C_{\lambda}^{\dagger}, \zeta\r]\r].}
We are interested in computing the elements of $\zeta$ which represent the
tunnelling probability between the vacua at $\tilde Y=0$ and $\tilde
Y=3\kappa$, so restricting our analysis to these off-diagonal elements, and
using the fact that in this basis $C_{\lambda}$ is diagonal with its
components given by the values of the superpotential at the critical points,
$\l(C_{\lambda}\r)_j{}^k = W(X_j) \de_j{}^k$, we find \eZAPP\ gives
\eqn\eLAPP{{1\ov4x}{d\ov dx}\l(x{d\ov dx}\zeta_{j\ol k}\r) = |W(X_j) -
W(X_k)|^2\zeta_{j\ol k}.}
Defining
\eqn\eMAPP{m_{jk} = 2|\lambda W(X_j) - \lambda W(X_k)|,}
and treating $\zeta$ as a function of this variable, gives the following
differential equation
\eqn\eNAPP{{d\ov dm}\l(m{d\ov dm}\zeta(m)\r) = m\zeta(m).}

The general solution, which vanishes in the large $\lambda$ limit, to this
differential equation is given by
\eqn\eGAPP{\zeta_{j\ol k} = C_{j\ol k}K_0(m_{jk}),}
where $C_{j\ol k}$ is some complex coefficient and $K_0(m)$ is given by
\eqn\eKAPP{K_0(m) = \int_{-\infty}^{\infty}{dp\ov 2\sqrt{p^2+m^2}}\exp\l(-
\sqrt{p^2+m^2}\r).}
In the limit $m\rightarrow\infty$, $K_0(m)$ has the asymptotic expansion
\eqn\eHAPP{K_0(m)\approx\sqrt{{\pi\ov2m}}\e^{-m}F(m),}
where
\eqn\eDEFF{F(m) = 1 + \sum_{l=1}^{\infty} {1\ov l!}[(2l-1)!!]^2(-8m)^{-l}.}
Since $\hbar\sim|\lambda|^{-1}$, the authors of \rCV\ have suggested that this
expansion may be interpreted as loop corrections to the one-instanton process.
In \figs{\fGot,\fGit}\ the first seven terms in the above expansion were used
to calculate the asymptotic behavior displayed.

Transforming this result for $\zeta$ back into the original point basis leads
directly to the results given in \eGoi\ and \eGoii.

\listrefs
\listfigs
\bye